\documentclass[prl,twocolumn,showpacs,floatfix]{revtex4}
\usepackage[T1]{fontenc}
\usepackage[cp1250]{inputenc}
\usepackage{psfrag}
\usepackage{color}
\usepackage{graphicx}
\usepackage[english]{babel}
\newcommand{\bq}{\begin{equation}}
\newcommand{\eq}{\end{equation}}
\newcommand{\ba}{\begin{eqnarray}}
\newcommand{\ea}{\end{eqnarray}}
\begin{document}
\title{A dynamical phase transition in a model for evolution with migration}
\author{Bart{\l}omiej Waclaw}
\author{Rosalind J. Allen}
\author{Martin R. Evans}
\affiliation{SUPA, School of Physics and Astronomy, University of Edinburgh, Mayfield Road, Edinburgh EH9 3JZ, United Kingdom}
\noindent
\begin{abstract}
Migration between different habitats is ubiquitous among biological
populations. In this Letter, we study a simple quasispecies model for
evolution in two different habitats, with different fitness
landscapes, coupled through one-way migration. Our model applies to
asexual, rapidly evolving organisms such as microbes. Our key finding is a dynamical phase transition at a critical value
of the migration rate. The time to reach steady state diverges at this
critical migration rate. 
Above the transition, the population is dominated by immigrants from the primary 
habitat. Below the transition, the genetic composition of the population 
is highly non-trivial, with multiple coexisting quasispecies which are not 
native to either habitat.
Using results from localization theory,
we show that the critical migration rate may be very small --
demonstrating that evolutionary outcomes can be very sensitive to even a
small amount of migration.
\end{abstract}
\maketitle

Biological dispersal---the movement of organisms between habitats---is
a ubiquitous phenomenon with important and wide-ranging consequences.
In the natural environment, organisms expand their ranges, colonise new habitats,
and can undergo speciation if they become spatially isolated.
Dispersal plays a key role in determining spatial and temporal patterns of genetic diversity in all organisms \cite{sakai}. 
For sexual organisms, with low mutation rates,  
population subdivision into demes, connected by migration, can have important effects on genetic diversity \cite{sexual,lande}, while in continuous space, transmission of unfit alleles can prevent the expansion of a species' range \cite{kirkpatrick}. 
For asexual, rapidly evolving organisms such as bacteria and viruses, dispersal also facilitates the emergence of new diseases and resistance to known treatments.
The ``source-sink'' paradigm \cite{holt,pulliam}, in which
migration from a favourable habitat maintains organisms in an
unfavourable one, has recently been used to explain the microbial genetics of
urinary tract infections \cite{sok}. However, despite its importance,
a general understanding of how migration  affects mutation-selection balance in microbial systems
is lacking.
In particular, one would like to know how migration changes the proportions of different genotypes in the evolving population.

In order to study the role of migration we 
introduce in this letter a simple statistical physics model for the
evolutionary dynamics of migrating asexual organisms.
Our model comprises  two
environmental habitats (with different fitness landscapes)
coupled by one-way migration of organisms from the primary to the
secondary habitat. Using a quasispecies approach, 
we find that the model undergoes a dynamical phase transition:
at a critical value of the migration
rate, the time to reach the steady state diverges. For sub-critical migration rates, 
the steady-state population in the secondary habitat is made up of the organisms ``native'' (best adapted)
to this habitat, as well as other, non-trivial, quasispecies, which
are not native to either habitat. Above the critical migration rate,
the native quasispecies in the secondary habitat is wiped out by
immigrants from the primary habitat. We use 
results from localization theory to gain insight into the transition
and to show 
that the critical migration rate is typically small, demonstrating
that even a small amount of migration can have an important effect on
evolutionary dynamics.

In our model, organisms have $M$ possible genotypes. $N_i$ and $n_i$ denote the
abundance (number density) of organisms with genotype $i$ in the primary and secondary habitat, respectively.
The
populations in the two habitats are thus described by the vectors
$\vec{N}=(N_1,\dots,N_M)$ and $\vec{n}=(n_1,\dots,n_M)$.  Organisms
migrate from the primary to the secondary habitat with rate
$k$. Within each habitat, mutations transform organism
$i$ to $j$ with rate $\gamma A_{ij}$, where $A_{ij}$ is a symmetric
adjacency matrix, to be discussed later. Organisms of type $i$
reproduce at a rate $\Phi_i-\sum_j N_j$ in the primary habitat and
$\phi_i-\sum_j n_j$ in the secondary habitat. The vectors
$\vec{\Phi}=(\Phi_1,\dots,\Phi_M)$ and
$\vec{\phi}=(\phi_1,\dots,\phi_M)$ thus describe the fitness
landscapes (or the maximal growth rate for organisms with genotype
$i$) in each habitat. The terms $-\sum_j N_j$
and $-\sum_j n_j$ in the growth rates account for population
saturation due to finite resources, as in the logistic equation. This
model is based on the para-mu-se (parallel mutation and selection)
\cite{baake} version of quasispecies theory \cite{eigen}, widely discussed in the biological, chemical and physical
literature
\cite{wilke,nowak,quasispecies,demetrius,leuthauser,baake2}.

The time evolution of the system is governed by the following set of equations for $i=1,\dots,M$:
\ba
	\dot{N}_i &=& N_i (\Phi_i-\sum_j N_j) + \gamma \sum_j A_{ij} (N_j- N_i), \label{eq1} \\
	\dot{n}_i &=& n_i (\phi_i-\sum_j n_j) + \gamma \sum_j A_{ij} (n_j- n_i)  + k N_i, \label{eq2}	
\ea where we have assumed that the primary habitat is large, so that
the loss of individuals due to migration has a negligible effect on
its population \cite{footnote}. For the calculations presented here,
we suppose that the fitness values $\Phi_i,\phi_i$ are independent
random numbers drawn from a distribution $P(\varphi)$, common to both environments. Thus genomes which are well-adapted in the
primary habitat are likely to be maladapted in the secondary habitat.

We first present the  analytical solution for the steady state \cite{WAE}.
For the primary habitat it is known from quasispecies theory \cite{baake,eigen}
that the steady-state abundances $\vec{N}^*$ are
\bq 
\vec{N}^* = \frac{\Lambda_1}{(\vec{\Psi_1}^T\cdot \vec{e})}
\vec{\Psi}_1, \label{vecnn} 
\eq 
with $\vec{e}=(1,\dots,1)$. Here $\Lambda_1$ is the largest eigenvalue of 
the matrix $W_{ij} = \delta_{ij} \Phi_i +
\gamma\Delta_{ij}, (\Delta_{ij}=A_{ij}-
\delta_{ij}\sum_k A_{ik }$ being the graph Laplacian),  and $\vec{\Psi}_1$ is the
corresponding eigenvector. 
We now determine the steady-state genotype abundances in the secondary habitat
by expanding in the eigenbasis of $V_{ij}=\delta_{ij} \phi_i + \gamma\Delta_{ij}$
\bq
	\vec{n}^*= k \sum_{\alpha =1}^{M} \frac{\vec{\psi_\alpha}^T\cdot \vec{N}^*}{n_{\rm tot} - \lambda_\alpha}\, \vec{\psi}_\alpha, \label{vecn}
\eq
where $\vec{\psi}_\alpha$ and $\lambda_\alpha$ are the eigenvectors and eigenvalues 
of $V_{ij}$ (ordered as $\lambda_1>\lambda_2>\dots$) 
and $n_{\rm tot}$ 
is the total steady-state population in the secondary habitat,
which is determined self-consistently as the largest root of
\bq
 n_{\rm tot} = k \sum_{\alpha=1}^M \frac{(\vec{\psi}_\alpha^T\cdot \vec{N}^*)(\vec{\psi}_\alpha^T\cdot \vec{e})}{n_{\rm tot} - \lambda_\alpha}.  \label{finfty}
\eq

\begin{figure}
\psfrag{k0}{$k=0$}\psfrag{k1}{$k=0.0001$}\psfrag{k2}{$k=0.001$}\psfrag{k3}{$k=0.002$}\psfrag{k4}{$k=0.003$}\psfrag{k5}{$k=1$}
\psfrag{i}{$i$}\psfrag{p}{$n_i^*$}\psfrag{p0}{$N_i^*$}
\includegraphics*[width=8cm]{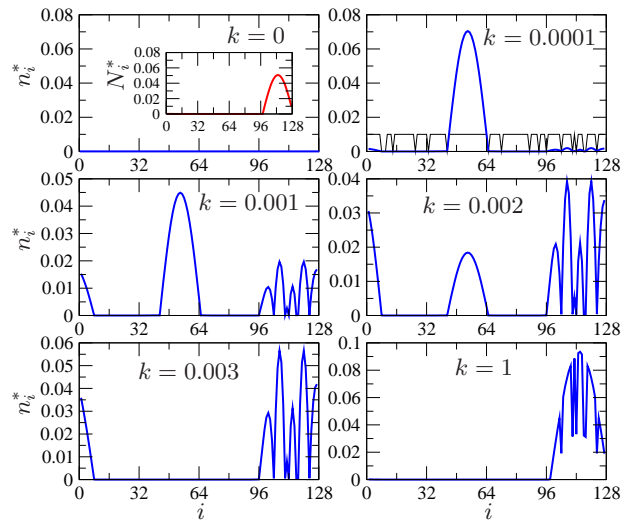}
\caption{Examples of steady-state genotype abundances in the secondary habitat $n_i^*$, for different values of the
migration rate $k$ where  $p=0.9$, $\gamma = 0.01$ and $M=128$. 
These results were obtained by numerical
self-consistent solution of Eqs.~(\ref{vecnn}), (\ref{vecn}) and
(\ref{finfty}). The top left panel shows zero abundances ($n_i^*=0$) in the absence of migration.
The inset shows the abundances $N_i^*$ in the primary habitat (red line). The top
right panel shows the ``native'' steady-state $n_i^*$ (blue line), for very small migration rate,
as well as the fitness landscape $\phi_i/100$ (black line). The other
panels show $n_i^*$ in the secondary habitat, for various values of the migration rate $k$. }
\label{f1}
\end{figure}

\begin{figure}
\psfrag{xx}{$k/k_0$} \psfrag{yy}{$T$} \psfrag{kk}{$k$}
\includegraphics*[width=\columnwidth]{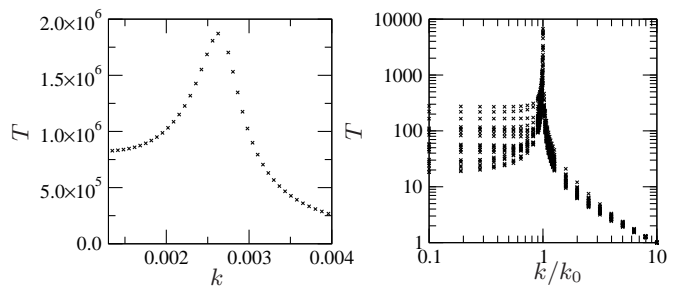}
\caption{Numerical results (from Eq.~(\ref{eq2}), using Eq.~(\ref{vecnn})
 for $N_i^*$)) for the time $T$ to reach the steady-state, starting
 from $N_i = N_i^*$ and $n_i = 0$. Left panel: $T$ as a function of
 migration rate $k$, for the same system as in Fig.~\ref{f1}.  The
 steady state was assumed to have been reached when $\sum_i
 |n_i(t+1)/n_i(t)-1|/M<10^{-10}$. Right panel: $T(k/k_0)$, where $k_0$
 is determined from Eq.~(\ref{k0}), for $M=64,\gamma=0.01,p=0.7$,
 normalized so that $T(10) = 1$. 
Results for 20 representative sets of
$\vec{\Phi},\vec{\phi}$ are presented on a log-log plot.}
\label{f4}
\end{figure}

To proceed further, we now make some specific assumptions about the
structure of the genome space (the mutation matrix $A_{ij}$) and the
fitness landscape $P(\varphi)$. To this end, we suppose that the
mutation graph is a one dimensional closed chain, in which mutations
are possible only between neighbouring genotypes ({\em{i.e.}}
$A_{ij}=1$ if $i=(M+j\pm 1){\rm mod} M$, and zero otherwise). We
further suppose that the fitness 
can take only two values: $1$ and $0$ with probability $p$ and $1-p$,
respectively. Since it has been suggested that viable genotypes form an interconnected network
in genome space \cite{gavrilets}, we shall consider the case $p\approx 1$, so
that the fitness landscape is characterised by ``islands'' of fit genotypes separated
by unfit ones.

Figure \ref{f1} shows how the population composition in the secondary
 habitat depends on the migration rate $k$. 
When $k$ is very small, the
 steady-state distribution $n_i^*$ is peaked around the
 longest sequence of maximal fitness values: this peak corresponds to
 the ``native'' (or best-adapted) quasispecies for the secondary
 habitat. When $k$ is very large (much larger than the mutation rate
 $\gamma$), the secondary habitat becomes dominated by immigrants from
 the primary habitat and the distribution of $n_i^*$ tends to the
primary-habitat
 steady-state distribution $N_i^*$.

\begin{figure}
\psfrag{k}{$k/k_0$} \psfrag{r}{$r/r_0$} \psfrag{f}{$n_{\rm tot}$}
\includegraphics*[width=\columnwidth]{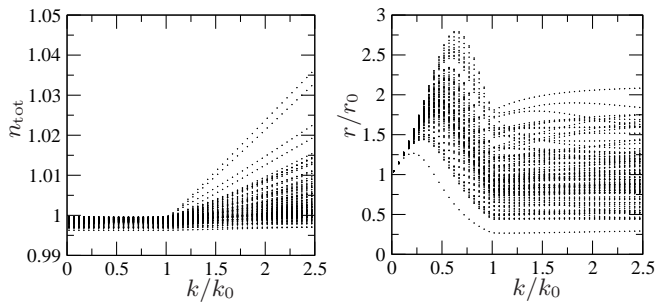}
\caption{Left: Total steady-state population $n_{\rm tot}$ in the
secondary habitat, as a function of rescaled migration rate $k/k_0$,
for $M=128$, $p=0.9$, $\gamma=0.1$, for a number of typical  random fitness
landscapes. 
Right: The normalized
participation ratio $r/r_0$ for $r_0=r(k\to 0)$, for the same set of
fitness landscapes.}
\label{f2}
\end{figure}

In contrast, for intermediate migration rates, the genetic composition
in the secondary habitat is highly non-trivial. As $k$ increases from
zero, the  quasispecies native to the secondary habitat is joined by
additional, non-native, quasispecies peaks. These do not correspond to
the native quasispecies from the primary habitat, but are instead
determined by the overlap of eigenvectors in the primary and secondary
habitats (as in Eq.~(\ref{vecn})). As the migration rate is increased
slightly further from 0.002 to 0.003, these new peaks dominate
completely and the native quasispecies of  the secondary
habitat disappears. 
This effect can be triggered by a very moderate change in
the migration rate. 
The appearance of these new quasispecies peaks suggests that
migration coupled to mutation can provide a mechanism for generation
and maintenance of genetic diversity (as will be shown later in Figure
\ref{f2}). Finally, as the migration rate increases further, the new peaks merge 
into the native quasispecies peak from the primary habitat.

\begin{figure}
\psfrag{t1}{$t=1$}\psfrag{t2}{$t=2^{5}$}\psfrag{t3}{$t=2^{10}$}\psfrag{t4}{$t=2^{7.5}$}\psfrag{t5}{$t=2^{15}$}\psfrag{t6}{$t=2^{16}$}
\psfrag{t7}{$t=2^{18}$}\psfrag{n}{$n_i$}\psfrag{i}{$i$}
\psfrag{k1}{$k=0.0001$}\psfrag{k2}{$k=0.003$}
\includegraphics*[width=8.0cm]{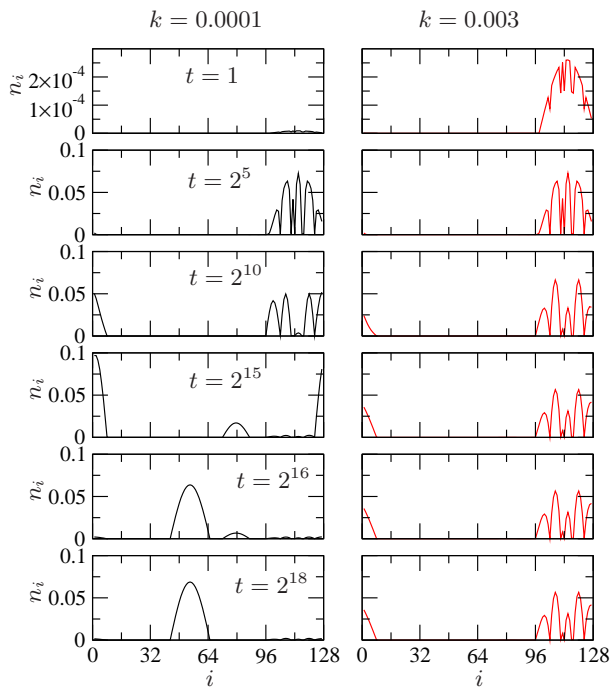}
\caption{Genetic composition in the secondary habitat $n_i(t)$, during the approach to steady state, for the same system as in Figure \ref{f1}, for two migration rates  $k=0.0001$ ($k \ll k_0$, black line, left) and $k=0.003$ ($k \gg k_0$, red, line right, the same vertical scale), for  $t=1,\dots,2^{18}$.}
\label{f3}
\end{figure}

Figure \ref{f4} (left panel) shows that this non-trivial dependence of
the steady state population on the migration rate is accompanied by
striking changes in the system dynamics. 
The time to reach the steady state plotted as a function of $k$
shows  a striking
maximum at $k_0\approx 0.0027$, suggesting a critical slowing down and
a likely dynamical phase transition. The approach to the steady state
for $k \ll k_0$ is much slower than for $k \gg k_0$. Figure \ref{f3}
illustrates the underlying reason for this. Here we plot snapshots of
$\vec{n}$ at various moments in time during the approach to the steady
state, for the same parameter set and fitness landscape, for migration
rates below and above $k_0$. For both migration rates, the immigrating
population initially has the same composition as the primary
habitat. For $k \ll k_0$, the primary habitat quasispecies
peak is lost entirely and the system undergoes  a slow process
of  jumps between
various local fitness maxima before finally settling in the global
optimum.  In contrast, for $k \gg
k_0$, the system 
rapidly relaxes to a steady state
which overlaps strongly with that of
the primary habitat.

Returning to our analytical expressions for $n_i^*$, Eqs.~(\ref{vecn})
and (\ref{finfty}), we can estimate the critical migration rate $k_0$
at which the dynamical phase transition takes place. Equation (\ref{vecn})
expresses $\vec{n}$ as a sum of eigenvectors $\vec{\psi}$ for the
secondary habitat, weighted by their overlap with $\vec{N^*}$. When $k
\to 0$, $n_{\rm tot}\to \lambda_1 \simeq 1$ and $\vec{n} \to
\vec{\psi}_1$ (see below). This is the native quasispecies
solution for the secondary habitat \cite{footnote1}. 
The phase transition occurs  when
this solution becomes dominated by the contributions from the other
terms ($\alpha=2,\dots,M$ in Eqs.~(\ref{vecn}) and (\ref{finfty})), which
arise from overlap with the primary habitat solutions. This happens at
a migration rate  approximately given by
\bq 
	k_0 = \lambda_1 \left(\sum_{\alpha=2}^M
	\frac{(\vec{\psi}_\alpha^T\cdot \vec{N}^*)(\vec{\psi}_\alpha^T\cdot \vec{e})}{\lambda_1 -
	\lambda_\alpha} \right)^{-1}. \label{k0} 
\eq 
To show that this result indeed
corresponds to the critical migration rate at which the transition happens, we plot in Figure~\ref{f4}, right panel,
the time $T$ to reach steady state as a function of $k/k_0$, where $k_0$ is determined from (\ref{k0}),
for simulated dynamics on $\approx 20$ representative random fitness
sequences $\vec{\Phi},\vec{\phi}$. Each fitness landscape generates a
slightly different curve $T(k/k_0)$, but all the curves appear to diverge
at $k=k_0$, indicating a phase transition. 

We now briefly discuss how the steady-state properties of the system
are affected by this phase transition. For 
$k<k_0$, the $\alpha=1$ terms in Eqs.~(\ref{vecn}) and (\ref{finfty})
dominate, while for $k>k_0$, the terms $\alpha=2,\dots,M$
dominate. This has important consequences for the total population
$n_{\rm tot}$ in the secondary habitat: for $k<k_0$, $n_{\rm tot}
\simeq 1$ \cite{footnote1}, while for $k>k_0$, $n_{\rm tot}$ grows with $k$. 
This prediction is confirmed numerically in
Fig.~\ref{f2}, left panel, for a number of randomly generated fitness
landscapes.

Another important steady-state property 
is genetic diversity. We noted from
Figure~\ref{f1} that multiple quasispecies peaks can coexist in steady
state for intermediate migration rates. Figure~\ref{f2}, right panel,
plots the participation ratio (PR) $r=(\sum_i n_{i}^*)^2/\sum_i
(n_{i}^{*})^2$, as a function of $k/k_0$, for several realisations of the
fitness landscape. The PR is a convenient measure of diversity, which
shows how many of the $n_i^*$'s are much larger than zero. Figure
\ref{f2} shows that $r$ reaches a maximal value at about $k\approx
0.25-0.75k_0$ and then decreases as $k$ approaches $k_0$; above $k_0$,
the diversity remains approximately constant.
Thus for weak migration the genetic diversity is  increased whereas
for strong migration it is washed out.

For the one-dimensional model considered here,  results from
localisation theory \cite{lif} allow us to estimate the  value 
of the critical migration rate $k_0$.  The eigenvector equation 
for $\vec{\Psi}_\alpha$ 
in  the primary
habitat
maps onto a Schr\"odinger
equation with  random potential $U_j=-\Phi_j/\gamma$: 
\bq 
	-(\Delta
	\vec{\Psi}_\alpha)_j + U_j (\vec{\Psi}_\alpha)_j = E_\alpha (\vec{\Psi}_\alpha)_j, 
	\label{SE}
\eq 
where $E_\alpha=-\Lambda_\alpha/\gamma$, and
likewise for an eigenvector $\psi_\alpha$ in the  secondary habitat. 
Equation (\ref{SE}) is essentially a 1D
tight-binding electron model \cite{ishii}, in which $U_j=-1/\gamma$
with probability $p$ 
and $U_j=0$ with probability $1-p$.  
Localization theory tells us that for
this problem the ground state eigenvector is localized,
taking the form $\Psi_{1,j} \sim \sin(j \pi/w)$ 
on the longest run $w$ of consecutive sites with
$U_j=-1/\gamma$, and  
has eigenvalue  $\Lambda_1 \simeq 1- \gamma \pi^2/w^2$.
Eigenvectors corresponding to excited states are similarly
localized on other, shorter potential wells.  To estimate $k_0$, we
observe that the largest contribution to the sum in (\ref{k0}) comes
from the eigenvector  with  the greatest overlap
with $\vec{N}^*$, which we denote $\vec{\psi}_\beta$. 
Assuming that $\vec{N}^*$ and $\vec{\psi}_\beta$ are
localized on potential wells of length
$w$ and $v$, respectively, 
we can estimate that $(\vec{\psi}_\beta^T \cdot \vec{N}^*)(\vec{\psi}_\beta^T \cdot \vec{e})\sim
v/w$. The lengths $w,v$ are the longest runs of $U_j=-1/\gamma$ in
sequences of independent binary random numbers of length
$M$ and $w$, respectively,  therefore $w\simeq \ln(M(1-p))/\ln(1/p)$ and
$v\simeq \ln(w(1-p))/\ln(1/p)$. For large $M$, $v$ is  much smaller
than $w$, so $\lambda_1-\lambda_\beta \simeq \gamma\pi^2/v^2$. Inserting
this into Eq.~(\ref{k0}), and setting
$\epsilon = 1-p$, we finally obtain
\bq \left<k_0\right> \sim
\gamma w/v^3  \simeq \frac{ \gamma \epsilon^2 \ln( M \epsilon)}{
(\ln \ln M\epsilon)^3}\;.
\label{k0approx1} \eq 
Remarkably, this rough estimate agrees up to a
factor $\approx 2$ with our simulation results.
Here  we have considered small $\epsilon$, where multiple fit genomes lie close
together in genotype space,  and we 
see from (\ref{k0approx1}) that $\left<k_0\right>$ is much
smaller than $\gamma$ for moderately large $M \epsilon $. This means that even a
very small migration rate (smaller than the mutation rate) can
dramatically change the course of evolution in the secondary
environment \cite{footnote2}.

In summary, we have introduced
a simple model for the evolution of asexual organisms in
two coupled habitats with different
fitness landscapes. We have shown that a dynamical phase transition occurs as the migration rate changes.
Bifurcations caused by migration have been observed in several models of sexual populations \cite{kirkpatrick,lande} but,
to our knowledge, the present work is the first to consider the effects of
migration on the evolutionary dynamics of asexual organisms from a 
quasispecies perspective.
In our model, at the critical migration rate, the population in the secondary
habitat becomes dominated by immigrants from the primary
habitat. 
For subcritical migration rates, our quasispecies model also reveals that migration can provide a novel mechanism for creation and maintenance of genetic diversity.

To obtain analytical results
and clear insights into the physics of the model, we have mainly considered a simple one-dimensional
closed-chain representation of the genotype space and binary random
fitness landscapes. As a step towards more complex and realistic representations of the genome space and fitness landscape, we
have also carried out numerical simulations for a continuous, uniform
distribution of the fitness, as well as a hypercubic mutation graph.
Our key results (in particular the
dynamical phase transition as a function of migration rate) remain
valid in these cases, suggesting that our findings are likely to be of
general significance. It will be interesting to extend our work to empirical fitness landscapes generated from experimental data \cite{real}, and, inspired by existing models for sexual organisms \cite{lande,kirkpatrick}, and recent models in microbial ecology \cite{warren}, to  multiple connected habitats and spatially varying environments.

\begin{acknowledgements}
The authors are grateful to N. H. Barton, R. A. Blythe, A. Free, and J.-M. Luck for valuable
discussions. R.J.A. was supported by the Royal Society. This work was funded by the EPSRC under grant EP/E030173.
\end{acknowledgements}


\begin{thebibliography}{100}
\bibitem{footnote} The inclusion of an extra term $-kN_i$ in Eq.~(\ref{eq1}) to account for the loss of organisms from the primary habitat due to migration would  be equivalent to adding a constant to the fitness landscape $\vec{\Phi}$ and would have no qualitative effect on the results reported here.
\bibitem{footnote1} 
$n_{\rm tot} \simeq 1$ is a consequence of our choice of constants
and does not imply a single organism.
For example, choosing $\Phi_i =c$ for fit phenotypes
yields $n_{\rm tot} \simeq c$
\bibitem{footnote2} One can perform a similar analysis for $p\ll 1$,  for which fit genomes are widely separated in genome space, and check that $\left<k_0\right> \sim \lambda_1 = \sqrt{1+4\gamma^2} - 2\gamma$.
The migration threshold is of order $1$ in this limit, but the conclusion that migrants inundate better adapted genotypes at some critical migration rate $k_0$ remains valid.
\end{thebibliography}
\end{document}